\documentclass[twocolumn,showpacs,amsmath,amssymb,prl]{revtex4}

\usepackage{graphicx}
\usepackage{dcolumn}
\usepackage{bm}
\begin{document}


\title{Effect of strain and electric
field on the electronic soft matter in manganite thin films}

\author{Tara Dhakal}
\author{Jacob Tosado}%
\author{Amlan Biswas}
\affiliation{Department of Physics, University of Florida,
Gainesville, FL 32611}

\date{\today}

\begin{abstract}
We have studied the effect of substrate-induced strain on the
properties of the hole-doped manganite
(La$_{1-y}$Pr$_{y}$)$_{0.67}$Ca$_{0.33}$MnO$_{3}$ ($y$ = 0.4, 0.5
and 0.6) in order to distinguish between the roles played by
long-range strain interactions and quenched atomic disorder in
forming the micrometer-scale phase separated state. We show that a
fluid phase separated (FPS) state is formed at intermediate
temperatures similar to the strain-liquid state in bulk compounds,
which can be converted to a metallic state by applying an external
electric field. In contrast to bulk compounds, at low temperatures a
strain stabilized ferromagnetic metallic (FMM) state is formed in
the $y$ = 0.4 and 0.5 samples. However, in the $y$ = 0.6 sample a
static phase separated (SPS) state is formed similar to the
strain-glass phase in bulk compounds. Hence, we show that long-range
strain interaction plays a dominant role in forming the
micrometer-scale phase separated state in manganite thin films.

\end{abstract}

 \maketitle

Multiphase coexistence in hole-doped manganites is a result of the
competition between phases of different electronic, magnetic and
structural orders ~\cite{ahn,dagotto}. This competition leads to
large changes in the physical properties of manganites due to small
perturbations e.g. colossal negative magnetoresistance (CMR). At low
temperatures the two competing phases are the ferromagnetic metallic
(FMM) and charge-ordered insulating (COI) phases. In manganites with
greater average $A$-site cation radii ($\langle r_A \rangle$) and
consequently a larger effective one-electron bandwidth ($W$) (e.g.
La$_{1-x}$Ca$_x$MnO$_3$, $0.2 < x < 0.5$), the pseudocubic FMM phase
is favored at low temperatures ~\cite{hwang}. When smaller ions such
as Pr are substituted at the $A$-site, $\langle r_A \rangle$ and $W$
are reduced. In these compounds the double-exchange mechanism is
suppressed and hence the pseudotetragonal (distorted) COI phase has
a comparable free energy to the FMM phase, resulting in micrometer
scale phase separation ~\cite{uehara}. It was shown that in the
presence of quenched disorder introduced by the ions of different
radii, the similarity of the free energies leads to coexistence of
the two competing phases ~\cite{dagotto}. However, the observation
of martensitic strain accommodation in manganites ~\cite{podzorov}
and fluid-like growth of the FMM phase observed in magnetic force
microscopy (MFM) images of phase separated manganites ~\cite{amlan},
suggests that the phases are not pinned. In fact, due to this
behavior the phase separated state in manganites has been described
as an ``electronic soft matter" state ~\cite{Milward,dagotto}. These
observations can be explained by an alternative model which shows
that the different crystal structures of the FMM and COI phases
generate long range strain interactions leading to an intrinsic
elastic energy landscape, which leads to micrometer scale phase
separation even without quenched disorder ~\cite{ahn}. To understand
the underlying mechanism for micrometer scale phase separation in
manganites, it is essential to distinguish between the roles played
by quenched disorder and long range strain interactions. We can then
devise protocols to manipulate this phase separated state with
external parameters such as strain, electric field, light etc., and
propose possible technical applications.

If long range strain interactions are the principal cause of phase
coexistence then it should be possible to control the elastic energy
landscape with substrate induced strain, which is known to
dramatically affect the nature of structural transitions in
materials such as shape memory alloys and ferroelectrics
~\cite{ferroelectric}. On the other hand, the effect of quenched
disorder can be estimated from the effect of isovalent substitution
of La-ions by Pr-ions. In this paper we report our results on the
effect of substrate induced strain and isovalent substitution in
thin films of the manganite
(La$_{1-y}$Pr$_y$)$_{0.67}$Ca$_{0.33}$MnO$_3$ (LPCMO), and compare
our results to bulk LPCMO. The $T-H$ phase diagram of bulk LPCMO
clearly shows two distinct types of phase separation (PS), a
strain-liquid (dynamic PS) and a strain-glass (frozen PS) regions
~\cite{sharma}. The strain-liquid phase shows large fluctuations in
resistivity and a slow relaxation of the magnetization
~\cite{sharma}. We show that in thin films of LPCMO, a fluid phase
separated (FPS) state is formed at intermediate temperatures similar
to the strain-liquid state in bulk materials. However, a strain
stabilized FMM phase is formed at low temperatures leading to a
sharper and larger drop in resistivity compared to bulk samples.
This strain stabilized FMM phase transforms to a static phase
separated (SPS) state when the Pr content is increased. The SPS
state is analogous to the strain-glass state in bulk LPCMO. An
external electric field transforms the FPS state to a metallic
state. However, there is negligible electric field effect once the
sample reaches the SPS state.

\begin{figure}[b]
\includegraphics[height=11.7cm]{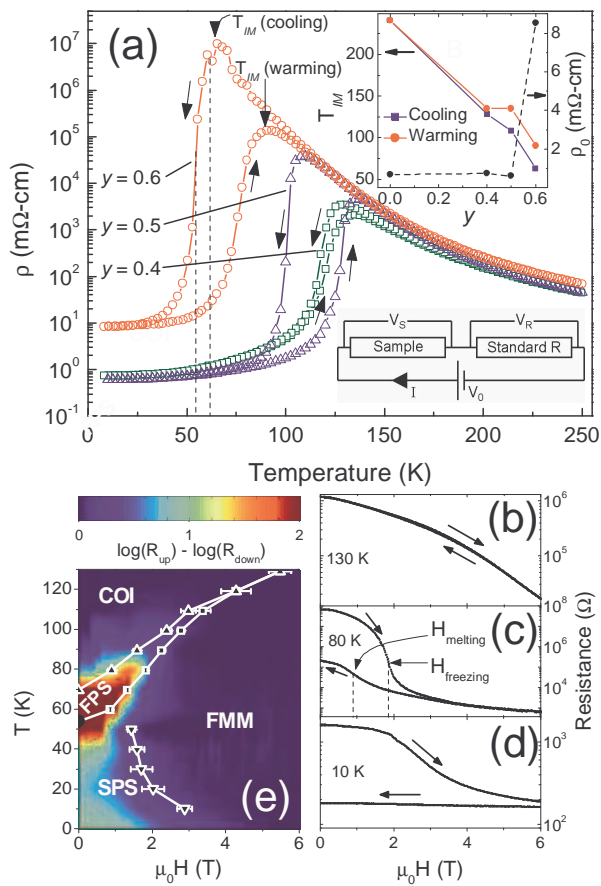}
\caption{(color online) (a) Resistivity vs. temperature curves for
thin films of (La$_{1-y}$Pr$_y$)$_{0.67}$Ca$_{0.33}$MnO$_3$ ($y$ =
0.4, 05, and 0.6). Cooling and warming directions are indicated by
arrows. The dotted lines mark the range of temperatures for the
$I-V$ curves shown in fig. 2. The upper inset shows the variation of
the transition temperatures and low temperature resistivity
($\rho_0$) with $y$. The lower inset shows the setup for measuring
the two probe resistance using a constant voltage source. (b), (c),
and (d) $R$ vs $H$ curves for the $y$ = 0.6 sample in the cooling
cycle. (e) The $T-H$ phase diagram for the $y$ = 0.6 thin film in
the cooling cycle.}
\end{figure}

We have grown thin films of
(La$_{1-y}$Pr$_y$)$_{0.67}$Ca$_{0.33}$MnO$_3$ (LPCMO) ($y$=0.4, 0.5
and 0.6) using pulsed laser deposition (PLD). The films were grown
in an oxygen atmosphere of 420 mTorr on NdGaO$_{3}$ (NGO) (110)
substrate kept at 820$^{\circ}$C. All the films described in this
letter are 30 nm thick and were grown at a rate of about 0.05 nm/s.
These growth conditions were optimized to obtain an
insulator-to-metal transition temperature while cooling, $T_{IM}$
(cooling), close to that observed in bulk compounds of similar
composition and the minimum transition width at $T_{IM}$ (cooling).
Such an optimization is crucial for mapping the phase diagram of
LPCMO, since the properties of thin films of this compound vary
markedly depending on the growth conditions. Standard
$\theta-2\theta$ x-ray diffraction data show that the films are
epitaxial and of a single chemical phase. Since the resistance of
the films can be as high as 1 G$\Omega$, the resistivity $\rho$, of
the films was measured with a two-probe method using a constant
voltage source, as shown in the lower inset of Fig. 1(a), with
$V_{0}$ set at 5 V ~\cite{resistance}. For the $\rho$ vs. $T$
curves, the temperature was varied at a rate of 2 K/min.

The $\rho$ vs. $T$ data for three LPCMO films ($y$ = 0.4, 0.5, and
0.6) are shown in Fig. 1(a). An expected reduction of $T_{IM}$
(cooling) is observed with increasing Pr concentration due to the
reduction of $\langle r_A \rangle$. The width of the hysteresis
between warming and cooling cycles of temperature drops sharply when
$y$ is reduced from 0.5 to 0.4 (Fig. 1(a), top inset). A remarkable
feature of the resistivity is that, unlike in bulk LPCMO
~\cite{uehara}, the residual resistivity $\rho_0$, (measured at 10
K) does not change from $y = 0$ (La$_{0.67}$Ca$_{0.33}$MnO$_3$,
LCMO) to $y = 0.5$ and then rises sharply for $y = 0.6$ (Fig. 1(a),
top inset). Resistivity measurements have been used to construct
temperature-magnetic field ($T-H$) phase diagrams, which elucidate
the stability of phases in manganites ~\cite{sharma}. To illustrate
the effect of substrate induced strain on the properties of LPCMO,
we mapped the $T-H$ phase diagram of the $y = 0.6$ sample by
measuring resistance $R$ vs. $H$ curves at different temperatures
for the cooling cycle (Fig. 1(b) to 1(d)). Since the effect of $H$
is irreversible, the sample was reset after every field sweep by
heating it to 150 K and then cooling it to the set temperature. The
data points were obtained by locating the magnetic field
corresponding to the steepest change in $R$ at a given temperature
i.e. where d$R$/d$H$ is maximum (Fig. 1(c)). The squares and
triangles represent the melting and freezing fields respectively
(Fig. 1(e)). The melting field line is extended to zero field by
including the $T_{IM}$ (cooling) at zero field from Fig. 1(a). The
inverted triangles represent the melting field at low temperature
($T \le$ 50 K). Since the transitions widths can be large in thin
films (Fig. 1(d)), as shown by the error bars in Fig. 1(e), we have
used a more direct method of constructing the $T-H$ phase diagram.
We plotted the difference in log($R$) between up sweep and down
sweep of $H$ log($R_{up}$)-log($R_{down}$), in the $T-H$ plane as a
2D color plot (Fig. 1(e)). The two methods of plotting the phase
diagram give similar results except for the SPS region due to the
broad transition with magnetic field at low temperatures. Four
distinct regions can be clearly identified in this phase diagram.
Two pure phases namely the COI state and the FMM state and two mixed
phase states namely, the fluid phase separated (FPS) state and the
static phase separated (SPS). The nomenclature of the mixed phase
states is based on the electric field effect, which will be
explained in detail in the following sections.

To elucidate the nature of phase coexistence in our thin films of
LPCMO we have to understand the combined effect of substrate induced
strain and Pr-substitution. Our LPCMO samples were grown on NGO
(110) substrates which are known to stabilize the pseudocubic FMM
phase in LCMO ~\cite{amlan}. Conversely, Pr has a smaller cation
radius than La and hence, substitution of La-ions by Pr-ions favors
a distorted crystal structure ~\cite{cox}. The distortion produced
by Pr-substitution reduces the $T_{IM}$ (cooling) in our LPCMO thin
films similar to bulk LPCMO ~\cite{uehara}. On the other hand,
strain induced by the NGO substrate removes the resistivity anomaly
at the charge-ordering temperature ($T_{CO}$) seen in bulk LPCMO
~\cite{uehara}, which is an effect of the suppression of the bulk
structural transition near $T_{CO}$ ~\cite{prellier}. It has been
shown for ferroelectric thin films that when the structural
transition is suppressed due to substrate strain, there is a strain
build-up in the film, which is released by the formation of domains
with different structures ~\cite{pompe}. A similar structural
separation in manganites would lead to phase separation into FMM and
COI regions and the observed reduction of $T_{IM}$ (cooling).
Furthermore, the thermal contraction of the substrate modifies the
strain landscape leading to the fluid nature of the phases as has
been observed in MFM images of LPCMO thin films
~\cite{biswasscience}. A detailed temperature dependent structural
study of the thin films is required to verify the above hypothesis.
However, it is clear from Fig. 1(a) and 1(e) that in spite of
substrate induced strain, the phase diagram of the LPCMO ($y$ = 0.6)
sample is similar to that of bulk LPCMO ~\cite{sharma}.

In contrast to the $y$ = 0.6 sample, $\rho_0$ for the $y$ = 0.4 and
0.5 samples drops to a value consistent with the pure FMM phase of
the $y$ = 0 sample (Fig. 1(a), upper inset). Magnetization
measurements also show that the $y = 0.5$ sample has a saturation
magnetization ($M_{sat}$) consistent with a pure FMM phase
~\cite{magnetization}. Hence, these two samples have a pure FMM
phase at low temperatures in contrast to the strain-glass phase
observed in bulk LPCMO ~\cite{sharma}, because the NGO substrate
favors the pseudocubic FMM phase at low temperatures. When the
Pr-concentration is increased to $y = 0.6$, $\rho_0$ increases by
about an order of magnitude to 8.7 m$\Omega$-cm. The value of
$M_{sat}$ for this sample is consistent with 50\% of the material
being in the FMM phase at low temperatures. As shown in Fig. 1(e), a
phase similar to the strain glass phase in bulk LPCMO (the SPS
state) appears in the phase diagram of thin film LPCMO only when the
Pr-concentration is increased to $y = 0.6$. It was suggested in ref.
~\cite{sharma} that the strain-liquid to strain-glass transition is
due to the interaction of the long-range strain with the quenched
atomic disorder. The observation of the FPS state and the absence of
the SPS state in the $y$ = 0.5 sample show that under substrate
induced strain, long-range strain is the driving force behind
micrometer scale phase separation in manganites.

To realize potential applications, an accessible handle is needed to
manipulate the phase separation in these thin films and one
candidate is an external electric field. Previous measurements of
the electric field effect showed a large drop in the resistivity of
charge-ordered manganites on the application of an electric field
~\cite{Asamitsu}. Electric field effects have also been observed in
thin films of LPCMO ~\cite{Pandey}. Large electric current effects
have been observed in LPCMO crystals due to Joule heating of the
metallic regions ~\cite{Tokunaga}.

\begin{figure}[t]
\includegraphics[height=7.7cm]{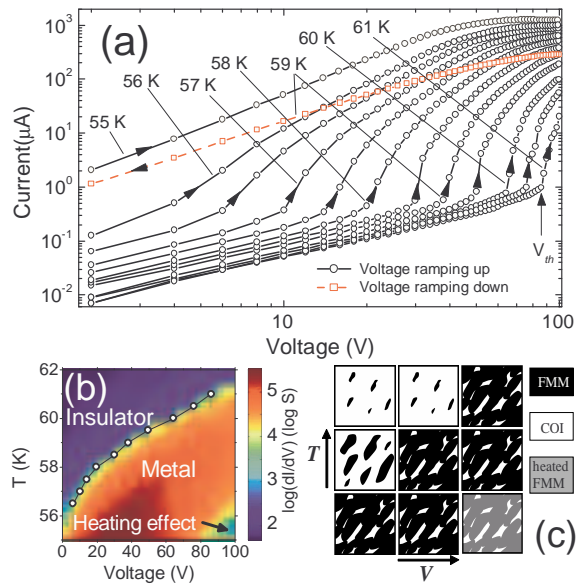}\\
\caption{(color online) (a) $I-V$ curves of the
(La$_{1-y}$Pr$_y$)$_{0.67}$Ca$_{0.33}$MnO$_3$ ($y$ = 0.6) thin film
in the cooling cycle with the voltage being ramped up. The dashed
curve is the $I-V$ curve at 59 K with the voltage being ramped down.
(b) The $T-V$ phase diagram for the $y$ = 0.6 thin film, showing the
variation of $V_{th}$ with temperature and the heating effect at
high voltages and low temperatures, in the cooling cycle. (c)
Schematic representation of the phase coexistence in the $T-V$
plane.}
\end{figure}

The voltage was applied to the LPCMO thin films using two indium
contacts 0.75 mm apart. The circuit for measuring the $I-V$ curves
is the same as the one used for measuring resistivity
~\cite{resistance}. Figure 2(a) shows the $I-V$ curves for the $y =
0.6$ sample for the cooling run. At a threshold voltage $V_{th}$,
the current across the film rises abruptly. This electric field
effect is irreversible as shown for the 59 K curve in figure 2a. The
sample stays in the low resistance state even when the electric
field is removed. To recover the high resistance state we heat the
film to a temperature above the resistivity hysteresis region and
then cool it down to the next desired temperature. We have plotted
the quantity log(d$I$/d$V$) calculated from these $I-V$ curves as a
function of temperature and voltage to construct a $T-V$ phase
diagram as shown in Fig. 2(b). The observed electric field effect is
not a heating effect as reported by Tokunaga \textit{et
al.}~\cite{Tokunaga} since heating should increase the resistance in
the temperature range shown in Fig. 2(a). We also observed a similar
electric field effect while cooling the $y = 0.5$ sample.

\begin{figure}[t]
\includegraphics[height=4.5cm]{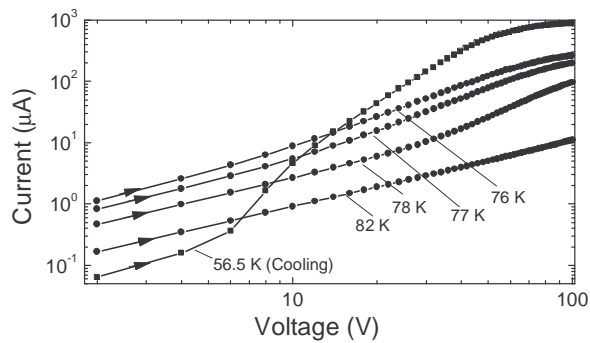}
\caption{$I-V$ curves of the
(La$_{1-y}$Pr$_y$)$_{0.67}$Ca$_{0.33}$MnO$_3$ ($y$ = 0.6) thin film
in the warming cycle with the voltage being ramped up. The $I-V$
curve at 56.5 K for the cooling cycle is shown for comparison.}
\end{figure}

A recent theoretical study using an extended double exchange
Hamiltonian showed that the application of an electric field favors
the FMM phase over the COI phase ~\cite{Gu}. The authors also
performed a numerical calculation using a random resistor network to
model a two-phase manganite sample, with low resistances
representing the metallic regions and high resistances for the
insulating regions. The electric field converts the high resistances
to low resistances and a percolation of the metallic regions leads
to the insulator to metal transition at $V_{th}$. $V_{th}$ decreases
as the number of high resistance elements is decreased from a 100\%
to 50\% of the sample, which qualitatively agrees with the observed
variation of $V_{th}$ as a function of temperature shown in Fig.
2(b) (decreasing temperature is analogous to decreasing number of
high resistances). A schematic picture of the phase coexistence in
LPCMO for the cooling cycle in the $T-V$ plane is shown in Fig.
2(c). As the size of the metallic regions (shown in black)
increases, the electric field across the smaller insulating regions
(shown in white) is enhanced, enhancing the local electric field
across the insulating regions. This enhancement of the local
electric field leads to the predicted and observed decrease in
$V_{th}$ with decreasing temperature. Above $V_{th}$, percolation of
the metallic regions in the film results in the sharp rise in the
conductivity of the film. Further rise in the voltage across the
film increases the current flowing through the metallic regions
resulting in local heating and a consequent decrease in conductivity
similar to the results of ref. ~\cite{Tokunaga}. This heating effect
is clearly seen in the bottom right corner of the $T-V$ phase
diagram in Fig. 2(b) (gray regions in Fig. 2(c)). When the sample is
cooled down to the SPS state, the metallic regions form a
percolating path. Hence, in the SPS state increasing the voltage
across the sample results in a larger current and Joule heating.

No sharp increase in current was observed during the warming run as
shown in Fig. 3. The $I-V$ behavior during cooling for a similar
resistance value is shown for comparison. In the warming run, the
absence of a $V_{th}$ suggests that there is no enhancement of the
local electric field. An explanation for this effect can be found in
MFM measurements, which have shown that the FMM regions are static
in the warming run ~\cite{biswasscience}. Therefore, once the sample
is cooled to the SPS region, the FMM and COI phases are locked in
space. Since the FMM regions percolate through the sample,
application of a voltage leads to Joule heating of the FMM regions.
On further warming the FMM regions homogeneously transform to a high
temperature insulating phase and hence there is no local enhancement
of the electric field.

In conclusion, substrate induced strain modifies the mechanism of
micrometer-scale phase separation in manganites. At intermediate
temperatures ($T \sim T_{IM}$ (cooling)), long-range strain
interactions lead to a fluid phase separated state analogous to the
strain-liquid phase in bulk LPCMO ~\cite{sharma}. However, below a
critical Pr concentration ($y \le 0.5$) the FPS state transforms to
a strain stabilized FMM phase at low temperatures, unlike the
strain-liquid to strain-glass transition in bulk LPCMO
~\cite{sharma}. A static phase separated state analogous to the
strain-glass phase in bulk LPCMO is observed at low temperatures
only when the Pr concentration (and hence the quenched atomic
disorder) is increased above a critical value ($y \ge 0.6$). An
external electric field provides an effective means to modify the
phase separation in manganites since it lowers the resistance of the
FPS state by two orders of magnitude due to a local electric field
enhancement. However, an electric field has negligible effect on the
SPS state. Further experiments using low temperature MFM are needed
to find the microscopic mechanism of the electric field effect.

\bibliography{cmrnote}

\begin{thebibliography}{10}
\bibitem{resistance}
We measured the voltage $V_{R}$ across the standard resistor $R$. The current
  in the circuit then was $V_{R}/R$ and the voltage across the sample $V_{S} =
  V_{0} - V_{R}$. Hence, the sample resistance $R_S = (V_0 - V_R) \times
  R/V_R$. As a check, the low temperature resistivity was also measured using a
  standard four-probe method. The $I-V$ curves were measured by varying
  $V_{0}$.
\bibitem{magnetization}
$M_{sat}$ is estimated by measuring the $M-H$ loops for the thin films and then
  subtracting the paramagnetic background due to the NGO substrate.
\end{thebibliography}

\newpage

\end{document}